# High electrical conduction of Sb square net in anti-ThCr$_2$Si$_2$ type La$_2$O$_2$Sb thin film grown by multilayer solid-phase epitaxy

Yuki Yamamoto,[a] Hideyuki Kawasoko,[a] and Tomoteru Fukumura*[a,b]

[a] Department of Chemistry, Graduate School of Science, Tohoku University, Sendai 980-8578, Japan.
[b] Advanced Institute for Materials Research and Core Research Cluster, Tohoku University, Sendai 980-8577, Japan.

* tomoteru.fukumura.e4@tohoku.ac.jp

Anti-ThCr$_2$Si$_2$ type $RE_2$O$_2$Sb ($RE$ = rare earth) with Sb square net has shown insulating conduction so far. Here we report the synthesis of La$_2$O$_2$Sb epitaxial thin films for the first time by multilayer solid-phase epitaxy. The valence state of Sb was about −2 evaluated from X-ray photoemission spectroscopy measurement, and the indirect band gap of 0.17 eV was observed. The La$_2$O$_2$Sb epitaxial thin film showed unexpectedly high electrical conduction as a narrow gap semiconductor, whose resistivity at room temperature was approximately ten-thousand-fold lower than that of La$_2$O$_2$Sb bulk polycrystal, attributed to increased carrier mobility probably due to suppressed Sb dimerization.

## Introduction

Recently, monatomic honeycomb nets like graphene have been extensively studied.[1–4] In layered compounds, various monatomic square nets such as Si, Sb, and Bi square nets are formed in ZrSiS, SrMnSb$_2$, and SrMnBi$_2$, respectively.[5–8] The square net layered compounds exhibited interesting electronic properties such as large magnetoresistance in ZrSiS and quantum Hall effect in EuMnBi$_2$,[9,10] and are expected to be topological materials owing to the unusual negative valence state.[11] For example, anti-ThCr$_2$Si$_2$ type $RE_2$O$_2$Pn ($RE$ = rare earth; $Pn$ = Sb, Bi) consists of alternative stacking of electrically conducting $Pn$ square nets and insulating $RE_2$O$_2$ layers, and is expected to show metallic conduction due to partially filled p orbital of negatively divalent $Pn$.[12,13] Among them, $RE_2$O$_2$Bi was metallic, and its superconductivity and high carrier mobility were induced by expanding interlayer distance between Bi square nets.[14–18] In contrast, $RE_2$O$_2$Sb was reported to be insulating,[19–21] being attributed to charge density wave state by Sb dimerization and/or Anderson localization by Sb disorder.[19,20,22] In order to realize electrically conducting Sb square net in $RE_2$O$_2$Sb, the Sb dimerization and/or the Sb disorder have to be suppressed by applying external stimuli such as epitaxial strain. In this study, we synthesized La$_2$O$_2$Sb epitaxial thin film for the first time by multilayer solid-phase epitaxy. The La$_2$O$_2$Sb epitaxial thin film showed ten-thousand-fold lower resistivity at room temperature than that of La$_2$O$_2$Sb polycrystal, probably due to suppressed Sb dimerization.

## Experimental

La$_2$O$_2$Sb epitaxial thin films were synthesized by applying multilayer solid-phase epitaxy (Ref. 23) as illustrated in Fig. 1. [La$_2$O$_3$/La/La$_2$O$_3$/Sb]$_2$ multilayer was deposited as a precursor on a MgO (001) substrate at room temperature in Ar gas atmosphere by using magnetron sputtering with Sb (99.9%), La (99.9%), and La$_2$O$_3$ (99.9%) targets. The layer thickness in the precursor was summarized in Table S1. The base and working pressures were approximately $1.0 \times 10^{-8}$ and $1.0 \times 10^{-2}$ Torr, respectively. Subsequently, the precursor was *in-situ* heated to various growth temperatures ($T_g$) of 650–1050 °C with the rate of 30 °C/min, kept at $T_g$ for 11 min, and then quenched to room temperature. It is noted that the $T_g$ was much lower than that (~1500 °C) of bulk synthesis.[19] The typical film thickness was 80 nm. The lifetime of the La$_2$O$_2$Sb epitaxial thin films was typically few days in air, hampering the transmission electron microscope measurement. The crystal structure was evaluated by X-ray diffraction (XRD) using Cu Kα radiation (D8 DISCOVER, Bruker AXS). The chemical composition was evaluated by scanning electron microscope equipped with energy dispersive X-ray spectroscopy (SEM-EDS; S-4300, Hitachi High Technologies). The electronic state was evaluated by X-ray photoemission spectroscopy (XPS; Quantum 2000, ULVAC-PHI) after surface cleaning by *in-situ* Ar sputtering. The optical properties were evaluated by using Fourier transform infrared and ultraviolet-visible-near-infrared spectrometers (FT/IR-6600 and V-770, JASCO). The electronic transport properties were evaluated by van der Pauw method with physical property measurement system (PPMS, Quantum Design).

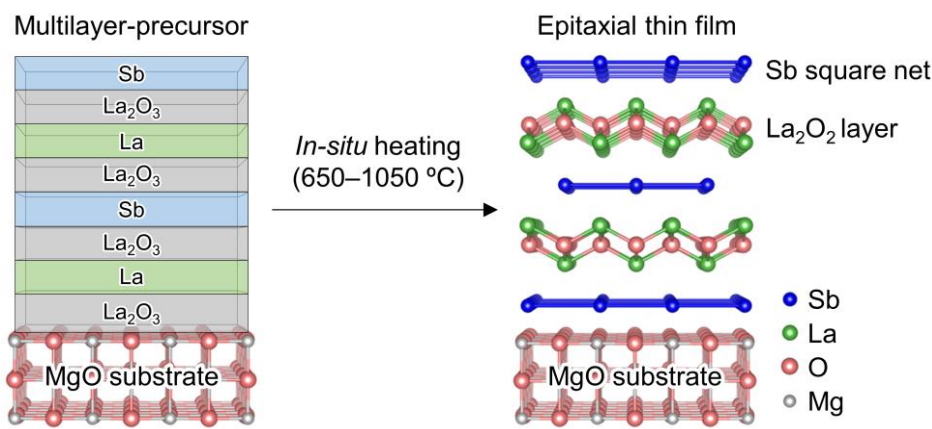

**Fig. 1** Schematic illustration of multilayer solid-phase epitaxy to synthesize $La_2O_2Sb$ epitaxial thin films. The crystal structures were drawn by the VESTA.[28]

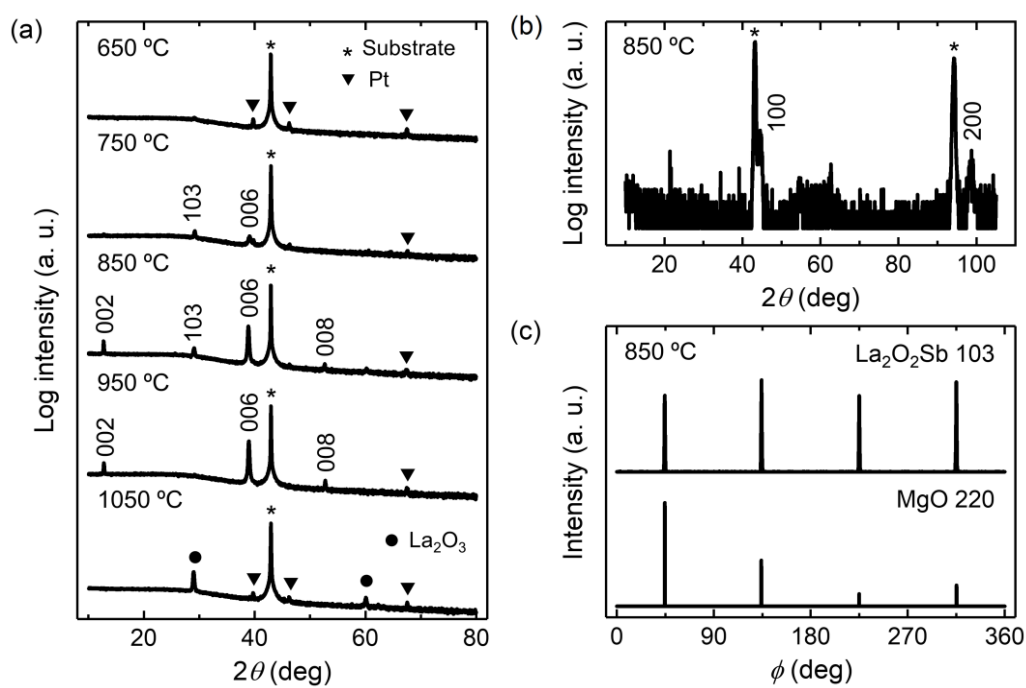

**Fig. 2** (a) Out-of-plane XRD patterns of the obtained films on MgO (001) substrates for various growth temperatures ($T_g$). (b) In-plane XRD pattern and (c) $\varphi$-scan of the La2O2Sb epitaxial thin film on MgO (001) substrate at $T_g$ = 850 °C.

## Results and discussion

Fig. 2a shows out-of-plane XRD patterns of the films grown at various $T_g$. At $T_g$ = 650 °C, XRD peaks of only MgO substrate and Pt paste attached on the substrate for thermal conduction were observed. At $T_g$ = 750 °C, $La_2O_2Sb$ 103 and 006 peaks were observed, indicating the crystallization of $La_2O_2Sb$ phase above 750 °C. With increasing $T_g$, the $La_2O_2Sb$ 00$l$ peaks were magnified, and only the $La_2O_2Sb$ 00$l$ peaks was observed at $T_g$ = 950 °C. Since film growth with $T_g$ = 950 °C sometimes resulted in the mixing of $La_2O_3$ phase (Fig. S2a), $T_g$ = 950 °C was an upper limit for the thin film growth due to the evaporation of Sb, although the $La_2O_2Sb$ thin films were reproducibly obtained with $T_g$ = 850 °C (Fig. S2b). At $T_g$ = 1050 °C, the $La_2O_2Sb$ peaks completely disappeared, while $La_2O_3$ phase appeared due to thermal vaporization of Sb as described below. The $La_2O_2Sb$ 200 and 400 peaks of grazing incidence in-plane XRD pattern for the film with $T_g$ = 850 °C showed the epitaxial growth of $La_2O_2Sb$ (001) thin films on MgO (001) substrate, representing the in-plane epitaxial relationship of $La_2O_2Sb$ [100] || MgO [100] (Fig. 2b). The absence of superlattice peak in the in-plane XRD pattern and the tetragonal symmetry measured by $\varphi$-scan (Fig. 2c) indicated negligible Sb dimerization in the Sb square net of the $La_2O_2Sb$ epitaxial thin films, in contrast with $Pr_2O_2Sb$ bulk single crystal.[21] Since the lattice constants were almost independent of the nominal amount of oxygen (Fig. S1a–S1c), different oxygen content would not influence the suppression of Sb dimerization. The reciprocal space mapping (RSM) of the $La_2O_2Sb$ epitaxial thin film (Fig. S1d) represented a relaxed growth of the $La_2O_2Sb$ epitaxial thin film on MgO substrate ($a$ = 4.21 Å). However, most of the $La_2O_2Sb$ epitaxial thin films showed longer $a$-axis length than $La_2O_2Sb$ bulk polycrystal (4.067 Å), suggesting that the epitaxial strain from MgO substrate suppressed the Sb dimerization in $La_2O_2Sb$ epitaxial thin films.

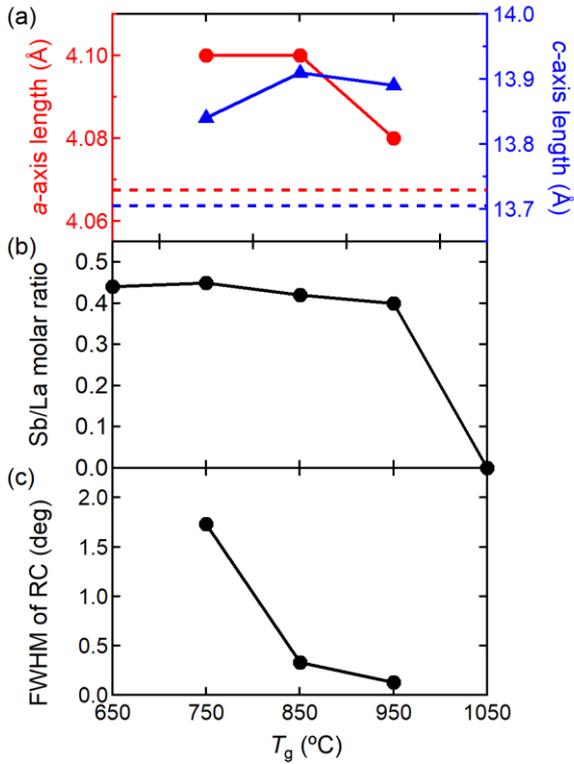

**Fig. 3** Growth temperature ($T_g$) dependence of (a) $a$- (circle) and $c$- axis lengths (triangle), (b) Sb/La molar ratio, and (c) FWHM of rocking curve (RC) for 006 diffraction peak of the $La_2O_2Sb$ epitaxial thin films. Red and blue dashed lines in (a) represent $a$- and $c$- axis lengths of $La_2O_2Sb$ bulk polycrystal, respectively.[19]

The lattice constants of the $La_2O_2Sb$ epitaxial thin films were $a$ = 4.08–4.10 Å and $c$ = 13.84–13.91 Å calculated from out-of-plane 006 and asymmetric 103 diffraction peaks. The $a$- and $c$-axis lengths of the $La_2O_2Sb$ epitaxial thin films were slightly longer than those of $La_2O_2Sb$ bulk polycrystal ($a$ = 4.0674 Å, $c$ = 13.705 Å),[19] probably because of the formation of oxygen vacancies in the films. Fig. 3b shows Sb/La molar ratio in the films as a function of $T_g$. Below $T_g$ = 950 °C, the Sb/La molar ratio was almost constant value of approximately 0.5, consistent with the formation of $La_2O_2Sb$ phase. The Sb/La molar ratio reduced from that of multilayer-precursors (Sb/La = 0.56) was caused by thermal vaporization of Sb during *in-situ* heating. At $T_g$ = 1050 °C, full thermal vaporization of Sb resulted in the formation of $La_2O_3$ phase (Fig. 2a). This full thermal evaporation is reasonable, taking into account the boiling temperature of Sb metal,[24]

550 °C at 1 × 10$^{-2}$ Torr. As shown in Fig. 3c, FWHM of rocking curve for the $La_2O_2Sb$ 006 diffraction peak was significantly narrowed down to be 0.13° with increasing $T_g$, indicating highly improved crystallinity of the $La_2O_2Sb$ epitaxial thin film.

Fig. 4 shows XPS spectra of Sb 3d, O 1s, and La 3d for the $La_2O_2Sb$ epitaxial thin film with $T_g$ = 950 °C. The Sb $3d_{3/2}$ and $3d_{5/2}$ peaks were located at 536.3 eV and 527.0 eV, respectively (Fig. 4a), which are lower in energy than those of Sb metal (537.4 eV for 3d$_{3/2}$, 528.0 eV for 3d$_{5/2}$),[25] indicating the negative valence state of Sb in the film. Also, the O 1s peak at 529.6 eV (Fig. 4a) and the two doublet peaks of La 3d$_{3/2}$ and 3d$_{5/2}$ at 852.3/855.5 eV and 835.4/838.5 eV, respectively (Fig. 4b), similar to those of La$_2$O$_3$ (529.4 eV for O 1s, 851.9/855.6 eV for La 3d$_{3/2}$, 835.1/838.8 eV for La 3d$_{5/2}$),[26,27] represented the formation of O$^{2-}$ and La$^{3+}$ in the film. These XPS results support negatively divalent Sb, trivalent La, and negatively divalent O in the film to fulfill the charge neutrality, corresponding to the partially filled Sb 5p orbital.[19,22]

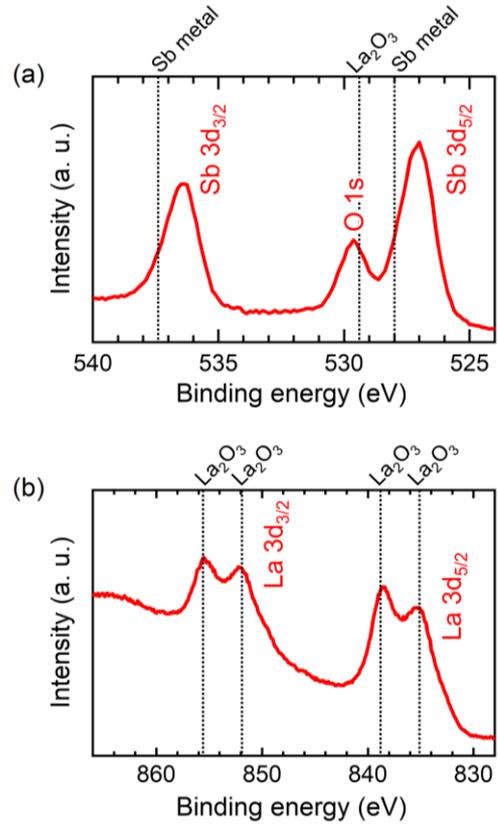

**Fig. 4** XPS spectra of (a) Sb 3d and O 1s and (b) La 3d peaks for the $La_2O_2Sb$ epitaxial thin film with $T_g$ = 950 °C.

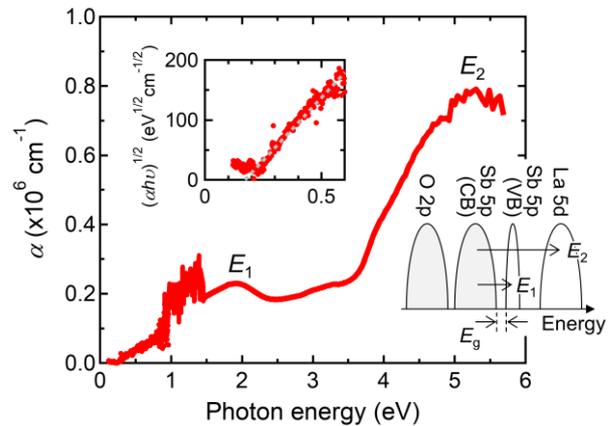

**Fig. 5** Absorption coefficient spectrum for the $La_2O_2Sb$ epitaxial thin film with $T_g$ = 850 °C. Upper inset shows Tauc plot of the absorption edge where the gray dashed line denotes the fitting line to evaluate the band gap. Bottom inset shows a schematic band structure of $La_2O_2Sb$.

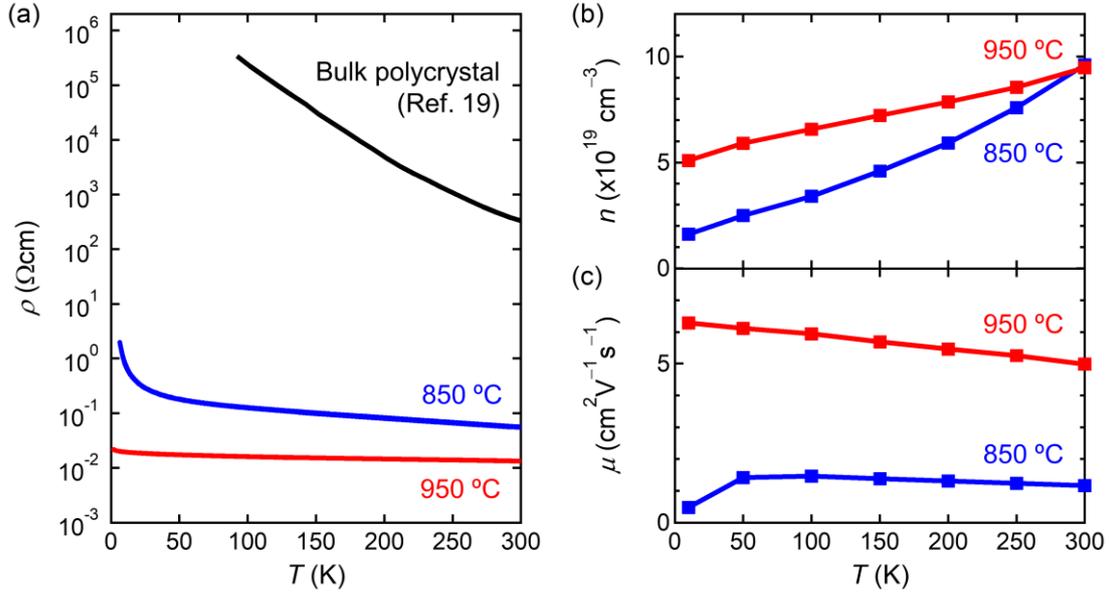

**Fig. 6** Temperature dependence of (a) resistivity, (b) carrier density, and (c) carrier mobility for the La$_2$O$_2$Sb epitaxial thin films with $T_g$ = 850 °C, 950 °C.

Fig. 5 shows the absorption spectrum for the La$_2$O$_2$Sb epitaxial thin film with $T_g$ = 850 °C. The Tauc plot of the absorption edge indicated the indirect band gap of 0.17 eV in the film (bottom inset of Fig. 5), corresponding to the energy splitting of Sb 5p orbital obtained by the band calculation, in which only Sb 5p orbital is located around Fermi energy.[19,22] This energy splitting of Sb 5p orbital would be attributed to slightly distorted herringbone-type Sb square net according to the first-principles calculation,[22] indicating suppressed Sb dimerization in the film. The absorption peaks at $E_1$ ~1.5 eV and $E_2$ ~5.2 eV would be attributed to electron transfers from Sb 5p valence band to Sb 5p conduction band and from Sb 5p valence band to La 5d, respectively, as schematically drawn in the upper inset of Fig. 5. Also, the almost identical optical absorption spectrum and bandgap were obtained for the different film with $T_g$ = 850 °C (Fig. S4).

Fig. 6 shows the electrical transport properties of the La$_2$O$_2$Sb epitaxial thin films with $T_g$ = 850 °C, 950 °C. In contrast with insulating La$_2$O$_2$Sb bulk polycrystal (3.2 × 10$^2$ Ωcm at 300 K),[19] the resistivity of the films (1.3–5.6 × 10$^{-2}$ Ωcm at 300 K) was approximately ten-thousand-fold smaller, in addition to the much weaker temperature dependence due to the narrow band gap. The films with $T_g$ = 950 °C showed lower resistivity than that with $T_g$ = 850 °C, being attributed to the improved crystallinity. The Hall coefficient showed linear magnetic field dependence with negative slope at all the temperatures (Fig. S5), indicating the n-type conduction probably caused by the oxygen vacancy, in contrast with the p-type conduction for the $RE_2$O$_2$Sb bulk polycrystals.[19] As shown in Fig. 6b, the films with $T_g$ = 850 °C, 950 °C showed the carrier density of 1.0 × 10$^{20}$ cm$^{-3}$ (300 K), slightly higher than 3–5 × 10$^{19}$ cm$^{-3}$ (300 K) for $RE_2$O$_2$Sb bulk polycrystal irrespective of $RE$ element.[19] In Fig. 6c, the carrier mobility of the films (1.0–5.0 cm$^2$V$^{-1}$s$^{-1}$ at 300 K) was much higher than that of La$_2$O$_2$Sb bulk polycrystal (≤ 10$^{-3}$ cm$^2$V$^{-1}$s$^{-1}$ at 300 K), probably due to suppressed Sb dimerization, in addition to the reduced grain boundary scattering. The film with $T_g$ = 950 °C showed the higher carrier mobility of 5.0–6.0 cm$^2$V$^{-1}$s$^{-1}$ than the film with $T_g$ = 850 °C, ~1.0 cm$^2$V$^{-1}$s$^{-1}$, indicating that the observed high carrier mobility was partially caused by the improved crystallinity. It is noted that resistivity, carrier density, and mobility of the films with $T_g$ = 850 °C were reproducibly obtained (Fig. S6).

## Conclusions

In summary, we synthesized La$_2$O$_2$Sb epitaxial thin films for the first time by multilayer solid-phase epitaxy. The La$_2$O$_2$Sb epitaxial thin film was a narrow gap semiconductor, exhibiting much higher electrical conduction than that of La$_2$O$_2$Sb bulk polycrystal, mainly because of the improved carrier mobility probably due to suppressed Sb dimerization. This result suggests that previously reported insulating nature of $RE_2$O$_2$Sb is attributed to not Anderson localization by Sb disorder but charge density wave state by Sb dimerization. Also, ideal Sb square net in $RE_2$O$_2$Sb would realize superconductivity and high carrier mobility as was observed in $RE_2$O$_2$Bi via oxygen intercalation.

## Acknowledgement

This study was supported by Yazaki Memorial Foundation for Science and Technology. YY acknowledges support from International Joint Graduate Program in Materials Science at Tohoku University.

## References

1. K. S. Novoselov, A. K. Geim, S. V. Morozov, D. Jiang, Y. Zhang, S. V. Dubonos, I. V. Grigorieva, A. A. Firsov, *Science* 2004, **306**, 666.
2. A. H. Castro Neto, F. Guinea, N. M. R. Peres, K. S. Novoselov, A. K. Geim, *Rev. Mod. Phys* 2009, **81**, 109.
3. A. J. Mannix, B. Kiraly, M. C. Hersam, N. P. Guisinger, *Nat. Rev. Chem.* 2017, **1**, 14.
4. N. R. Glavin, R. Rao, V. Varshney, E. Bianco, A. Apte, A. Roy, E. Ringe, P. M. Ajayan, *Adv. Mater.* 2020, **32**, 1904302.
5. L. Schoop, M. Ali, C. Straßer, A. Topp, A. Varykhalov, D. Marchenko, V. Duppel, S. Parkin, B. Lotsch, C. Ast, *Nat. Commun.* 2016, **7**, 11696.
6. J. Y. Liu, J. Hu, Q. Zhang, D. Graf, H. B. Cao, S. M. A. Radmanesh, D. J. Adams, Y. L. Zhu, G. F. Cheng, X. Liu, W. A. Phelan, J. Wei, M. Jaime, F. Balakirev, D. A. Tennant, J. F. DiTusa, I. Chiorescu, L. Spinu, Z. Q. Mao, *Nat. Mater.* 2017, **16**, 905.
7. K. A. Benavides, I. W. H. Oswald, J. Y. Chan, *Acc. Chem. Res.* 2018, **51**, 12.
8. J. Park, G. Lee, F. Wolff-Fabris, Y. Y. Koh, M. J. Eom, Y. K. Kim, M. A. Farhan, Y. J. Jo, C. Kim, J. H. Shim, J. S. Kim, *Phys. Rev. Lett.* 2011, **107**, 126402.
9. M. N. Ali, L. M. Schoop, C. Garg, J. M. Lippmann, E. Lara, B. Lotsch, S. S. Parkin, *Sci. Adv.* 2016, **2**, e1601742.
10. H. Masuda, H. Sakai, M. Tokunaga, Y. Yamasaki, A. Miyake, J. Shiogai, S. Nakamura, S. Awaji, A. Tsukazaki, H. Nakao, Y. Murakami, T. H. Arima, Y. Tokura, S. Ishiwata, *Sci. Adv.* 2016, **2**, e1501117.
11. S. Klemenz, A. K. Hay, S. M. L. Teicher, A. Topp, J. Cano, L. M. Schoop, *J. Am. Chem. Soc.* 2020, **142**, 6350.
12. B. Benz, *Acta Cryst.* 1971, **B27**, 853
13. J. Nuss, M. Jansen, *J. Alloys Compd.* 2009, **480**, 57.
14. H. Mizoguchi, H. Hosono, *J. Am. Chem. Soc.* 2011, **133**, 2394.
15. R. Sei, S. Kitani, T. Fukumura, H. Kawaji, T. Hasegawa, *J. Am. Chem. Soc.* 2016, **138**, 11085.
16. K. Terakado, R. Sei, H. Kawasoko, T. Koretsune, D. Oka, T. Hasegawa, T. Fukumura, *Inorg. Chem.* **2018**, *57*, 10587.
17. R. Sei, H. Kawasoko, K. Matsumoto, M. Arimitsu, K. Terakado, D. Oka, S. Fukuda, N. Kimura, H. Kasai, E. Nishibori, K. Ohoyama, A. Hoshikawa, T. Ishigaki, T. Hasegawa, T. Fukumura, *Dalton Trans.* 2020, **49**, 3321.
18. K. Matsumoto, H. Kawasoko, H. Kasai, E. Nishibori, T. Fukumura, *Appl. Phys. Lett.* 2020, **116**, 191901.
19. P. L. Wang, T. Kolodiazhnyi, J. Yao, Y. Mozharivskyj, *J. Am. Chem. Soc.* 2012, **134**, 1426.
20. P. L. Wang, T. Kolodiazhnyi, J. Yao, Y. Mozharivskyj, *Chem. Mater.* 2013, **25**, 699.
21. O. V. Magdysyuk, J. Nuss, M. Jansen, *Acta Cryst.* 2013, **B69**, 547.
22. H. Kim, C.-J. Kang, K. Kim, J. H. Shim, B. I. Min, *Phys. Rev. B* 2015, **91**, 165130.
23. R. Sei, T. Fukumura, T. Hasegawa, *ACS Appl. Mater. Interfaces* 2015, **7**, 24998.
24. G. M. Rosenblatt and C. E. Birchenall, *J. Chem. Phys.* 1961, **35**, 788.
25. M. Pessa, A. Vuoristo, M. Vulli, S. Aksela, J. Väyrynen, T. Rantala, H. Aksela, *Phys. Rev. B* 1979, **20**, 3115.
26. O. P. Ivanova, A. V. Naumkin, L. A. Vasilyev, *Vacuum* 1996, **47**, 67.
27. M. F. Sunding, K. Hadidi, S. Diplas, O. M. Løvvik, T. E. Norby, A. E. Gunnæs, *J. Electron Spectrosc. Relat. Phenom.* 2011, **184**, 399.
28. K. Momma, F. Izumi, *J. Appl. Crystallogr.* 2011, **44**, 1272.